%% file: main.tex
\documentclass[conference]{IEEEtran}
\IEEEoverridecommandlockouts
\usepackage{cite}
\usepackage{amsmath,amssymb,amsfonts}
\usepackage{algorithmic}
\usepackage{siunitx}
\usepackage{graphicx}
\usepackage{textcomp}
\usepackage{xcolor}
\usepackage[a4paper, total={184mm,239mm}]{geometry}
\def\BibTeX{{\rm B\kern-.05em{\sc i\kern-.025em b}\kern-.08em
        T\kern-.1667em\lower.7ex\hbox{E}\kern-.125emX}}

\newcommand{\respond}[1]{#1}

\newcommand{\systemname}{EmuNoC}
\newcommand{\eg}{\emph{e.g.,}}
\newcommand{\ie}{\emph{i.e.,}}
\usepackage{fancyhdr}
\usepackage[all]{background}
\usepackage{stackengine}
\setstackEOL{\\}
\setstackgap{L}{\normalbaselineskip}
\SetBgContents{\color{gray}{\tiny \Longstack{PREPRINT - Accepted in Proceedings of the 32nd International Conference on Field Programmable Logic (FPL '22)}}}
\SetBgPosition{4.2cm,1cm}
\SetBgOpacity{1.0}
\SetBgAngle{0}
\SetBgScale{1.8}

\usepackage{pgfplots}
\usepackage{pgfplotstable}
\usetikzlibrary{patterns, arrows, pgfplots.groupplots,hobby}
\usetikzlibrary{patterns}
\usetikzlibrary{positioning}
\usetikzlibrary{pgfplots.groupplots}
\usetikzlibrary{plotmarks}
\usetikzlibrary{calc}
\usepgfplotslibrary{external}
\usepackage{balance}
\usepackage{xcolor}
\usepackage{colortbl}
\usepackage{rotating}
\usepackage{multirow}
\usepackage{booktabs}
\usepackage{subfigure}
\usepackage{bm}
\usepackage{subfig}
\usepackage{enumitem}
\usepackage{soul}
\usepackage{listings}
\definecolor{cg1}{HTML}{E8F1FA}
\definecolor{cg2}{HTML}{C7DDF2}
\definecolor{cg3}{HTML}{8EBAE5}
\definecolor{cg4}{HTML}{407FB7}
\definecolor{cg5}{HTML}{00549F}
\definecolor{red}{HTML}{CC071E}
\input{figs/defs}

\usepackage{xcolor}
\definecolor{white}{rgb}{1.0, 1.0, 1.0}
\definecolor{black}{rgb}{0.0, 0.0, 0.0}
\definecolor{applegreen}{rgb}{0.55, 0.71, 0.0}
\definecolor{blue(pigment)}{rgb}{0.2, 0.2, 0.6}
\definecolor{turquoise}{rgb}{0.19, 0.84, 0.78}
\definecolor{sinopia}{rgb}{0.8, 0.25, 0.04}

\lstdefinestyle{CStyle}{
    language=C,
    backgroundcolor=\color{white},
    commentstyle=\color{applegreen},
    keywordstyle=\color{blue(pigment)},
    stringstyle=\color{sinopia},
    basicstyle=\tiny\ttfamily,
    breakatwhitespace=false,
    breaklines=true,
    keepspaces=true,
    frame=single,
    numberstyle=\tiny\color{black},
    numbers=left,
    numbersep=5pt,
    captionpos=t,
    showspaces=false,
    showstringspaces=false,
    showtabs=false,
    tabsize=2,
    morekeywords=[2]{uint8_t,uint16_t,uint32_t,uint64_t,int8_t,int16_t,int32_t,int64_t,string,vector,ustring,size_t,Elf64_Ehdr,Elf64_Shdr,Elf64_Sym,Elf64_Rela,Elf64_Obj,Elf64_ShdrTbl,Elf64_GenShdr,Elf64_SymTbl,Elf64_RelaTbl,Elf64_Progbits,ENCODE_TYPE,FILE,Info,TextFile},
    keywordstyle=[2]\color{turquoise},
    sensitive=true
}

\usepackage{amssymb}
\usepackage{pifont}

\usepackage{keyval}

\begin{document}
\bstctlcite{IEEEexample:BSTcontrol}

\title{
\systemname: Hybrid Emulation for Fast and Flexible Network-on-Chip Prototyping on FPGAs
\thanks{Funded by the Federal Ministry of Education and Research (BMBF) and the Ministry of Culture and Science of the German State of North Rhine-Westphalia (MKW) under the Excellence Strategy of the Federal Government and the Länder (G:(DE-82)EXS-SF-Project No. StUpPD\_390-21)}
}

\author{
    \IEEEauthorblockN{Yee Yang Tan\IEEEauthorrefmark{1}, Felix Staudigl\IEEEauthorrefmark{1}, Lukas J\"unger\IEEEauthorrefmark{1}, Anna Drewes\IEEEauthorrefmark{2}, Rainer Leupers\IEEEauthorrefmark{1}, and Jan Moritz Joseph\IEEEauthorrefmark{1}
                             }
    \IEEEauthorblockA{\IEEEauthorrefmark{1}\textit{Institute for Communication Technologies and Embedded Systems, RWTH Aachen University, Germany}\\
    \{tan, staudigl, juenger, leupers, joseph\}@ice.rwth-aachen.de\\
    }
    \IEEEauthorblockA{\IEEEauthorrefmark{2}\textit{Institute for Information and Communication Technologies, Otto-von-Guericke Universität Magdeburg, Germany}\\
    anna.drewes@ovgu.de\\
    }
}

\maketitle

\begin{abstract}
\input{abstract}
\end{abstract}

\begin{IEEEkeywords}
Hybrid emulation, NoCs, FPGA
\end{IEEEkeywords}


\section{Introduction}
\input{intro}

\section{Background and Related Works}\label{sec:back}
\input{background}

\section{Emulation System Architecture}
\label{sec:architecture}
\input{method}

\section{Results}
\label{sec:results}
\input{result}

\section{Conclusion}
\label{sec:con}
\input{con}

\IEEEtriggeratref{28} 
\bibliographystyle{IEEEtran}
\bibliography{biblio.bib}

\end{document}

%% file: figs/defs.tex
\tikzset{router/.style={circle,draw,fill=gray!60,inner sep=0pt,minimum size=5pt}}
\usetikzlibrary{fadings}
\tikzset{desc/.style={font = \footnotesize\sffamily}}

\definecolor{col1}{RGB}{27,158,119}
\definecolor{col2}{RGB}{217,95,2}
\definecolor{col3}{RGB}{117,112,179}


%% file: abstract.tex
Networks-on-Chips (NoCs) recently became widely used, from multi-core CPUs to edge-AI accelerators. Emulation on FPGAs promises to accelerate their RTL modeling compared to slow simulations. However, realistic test stimuli are challenging to generate in hardware for diverse applications. In other words, both a fast and flexible design framework is required. The most promising solution is hybrid emulation, in which parts of the design are simulated in software, and the other parts are emulated in hardware. This paper proposes a novel hybrid emulation framework called \systemname. We introduce a clock-synchronization method and \respond{software-only packet generation} that improves the emulation speed by 36.3$\times$ to 79.3$\times$ over state-of-the-art frameworks while retaining the flexibility of a pure-software interface for stimuli simulation. We also increased the area efficiency to model up to an NoC with 169 routers on a single FPGA, while previous frameworks only achieved 64 routers.

%% file: intro.tex
Today, massively-parallel multi-processors can be found in different forms in nearly any system. Their application ranges from conventional multi-core CPUs, e.g., in cloud servers, to massively scaled systems used as low-power neuromorphic edge AI accelerators. In any case, Networks-on-Chip (NoCs) have become the \respond{de facto} communication infrastructure for their excellent scaling capability. Examples showing the nearly universal use of NoCs are data-flow or parallel processors, e.g., manycores \cite{bakhoda2010throughput}, big data \cite{blochwitz2015optimized}, server-scale AI \cite{loihi}, edge AI \cite{neuronflow}, data bases \cite{blochwitz2017hardware}, in-memory computing \cite{guirado2022wireless}, genome sequencing~\cite{sarkar2009network},  medical applications \cite{passaretti2019survey}. 

Due to the wide use of NoCs, many design tools have been created, e.g.,  \cite{rataoskr-tomacs, noxim}. They are used for design space exploration (DSE) with software simulations or hardware prototypes that evaluate key performance metrics (KPIs) and guide the architect. Traditionally, NoC simulators target multi-core CPUs. With the emergence of edge AI accelerators, there is a need for more versatile tools for changing applications or mappings, as contributed by this work.

A typical DSE flow comprises the following: Architectural simulators are fast and flexible, providing early critical insights. After implementing a register transfer level (RTL) model, these can be simulated. This is slow and practically only helpful in verifying or evaluating small parts of any design. Therefore, emulation on FPGAs is highly relevant as of massive speed improvements \cite{drewes2017fpga, chu.2020} for any architecture optimization \cite{joseph2019nocs}. 

\begin{figure}
    \centering
    \includegraphics[width=1\linewidth]{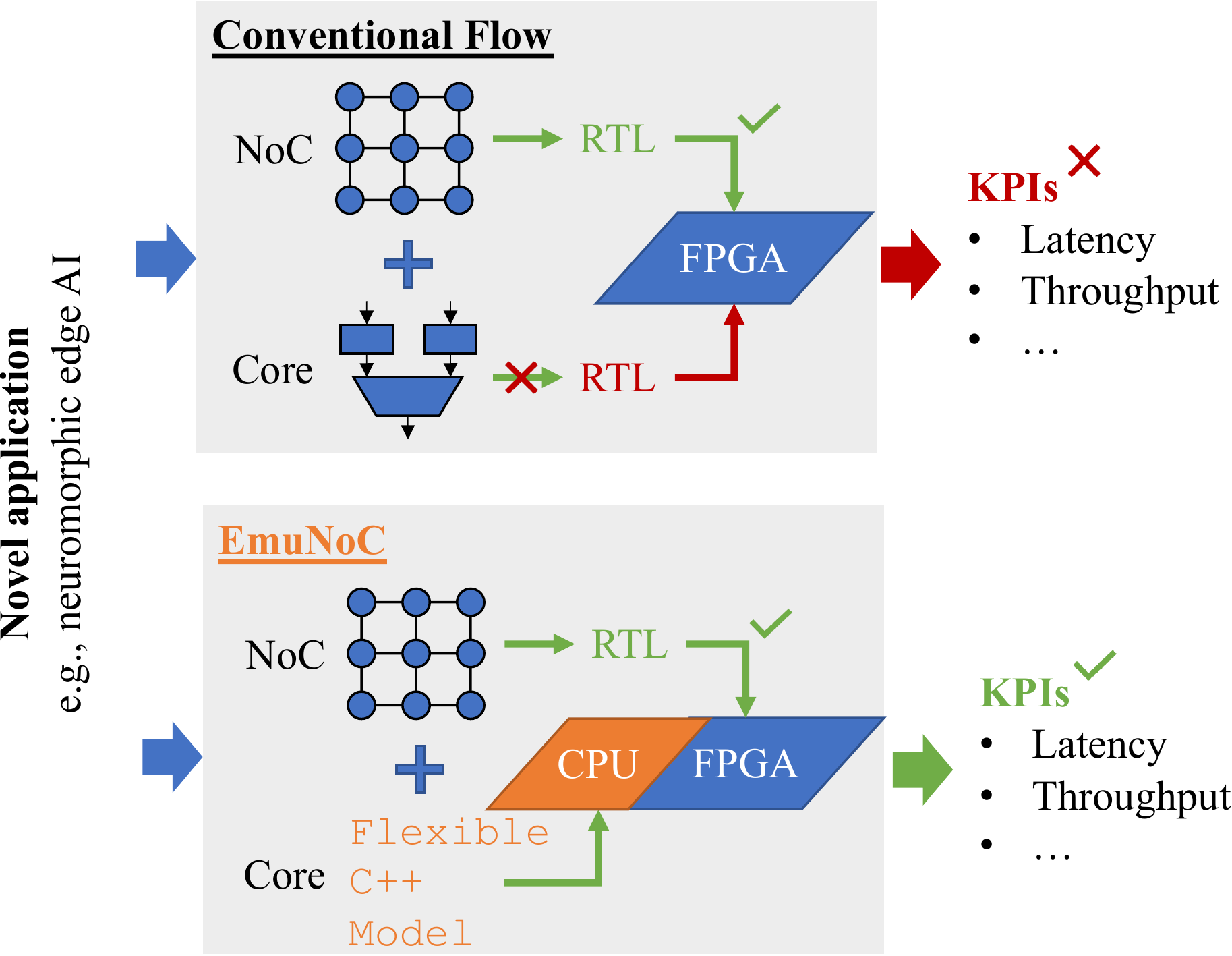}
    \caption{\systemname: Toolflow}
    \label{fig:intro}
\end{figure}

One key downside of FPGA emulation is its limited flexibility to adapt to novel use cases \cite{chu.2020}. In other words, for every new NoC use case, the RTL design of all cores, etc., would be required. The availability of these models is unrealistic in an early design stage, hindering the deployment of NoCs for emerging use cases. 

This paper provides an effective solution. We contribute an open-source framework\footnote{ \texttt{https://github.com/ICE-RWTH/EmuNoC}} for NoC hybrid emulation, in which the traffic pattern can easily be switched by software models, but the NoC is emulated on the FPGA for high performance and accuracy (Fig. \ref{fig:intro}). We propose a novel clock-synchronization method and hardware-only packet generation that improves the emulation speed by more than one magnitude. Specifically, this paper yields the following novelties:

\begin{itemize}
    \item 
We propose a hardware clock halting technique for faster hybrid emulation, which shows a speedup of 79.3$\times$ for synthetic traffic and 36.3$\times$ for Netrace \cite{netrace2010} over the previous state-of-the-art emulation framework.
    \item 
We contribute a novel concept of a single clock synchronous serializer used as a Network Interface (NI) for NoCs with virtual channels (VCs).
    \item We evaluate our system for different case studies (multi-core CPUs, edge AI accelerators) to show flexibility.
\end{itemize}

The paper is organized as follows. In Sec.~\ref{sec:back} we will introduce the background for hybrid emulation of NoCs and discuss the related works. In Sec.~\ref{sec:architecture}, we will explain our architecture. In Sec.~\ref{sec:results} we will analyze the system performance. Finally, the paper is concluded.

%% file: background.tex
\respond{When building efficient NoCs, they must be evaluated against benchmarks. For this, stimuli are to be generated. Synthetic traffic enables system validation, e.g., through fuzzy testing using random traffic, but it is not helpful for architectural exploration as it does not reflect real workloads. Full-system simulators (FSS), such as gem5, execute the whole system, including the operating system, cores, caches, and the NoC. The method offers the highest-precision benchmarks but often is unacceptably slow. Traces, recorded with an FSS and replayed later, provide a useful middle-ground.}

For modeling the NoC, the accuracy of traces is often sufficient, as demonstrated by Netrace \cite{netrace2010} for multi-core CPUs. Netrace provides trace files and a dependency-driven C-based player. The traces are generated by running PARSEC benchmarks on gem5 for a 64-core system. The dependency tracking between packets boosts the accuracy. For AI systems, trace generation is often more straightforward than for CPUs, because of the deterministic execution of DNNs (e.g., no situational caching). Most mapping strategies of DNNs enable mathematical trace modeling. One example is NewroMap \cite{newromap}, which showed that the feed-forward characteristic of DNNs can be exploited for mapping neurons to neuromorphic, memory-bound accelerators. The resulting traffic patterns in the NoC yield high locality and a low number of dependencies.

\begin{table}
    \centering
    \caption{NoC-on-FPGA emulators.}
    \resizebox{0.5\textwidth}{!}{\input{./tables/cmp.tex}}
    \label{tab:cmp_emu}
\end{table}

Cycle-accurate simulators, RTL simulation, or FPGA hybrid emulation offer different options to understand the performance of NoCs before their deployment to a system, hence enabling design space exploration.

Cycle-accurate simulators offer a decent compromise between accuracy and speed. Booksim \cite{booksim}, Noxim \cite{noxim}, and Ratatoskr \cite{Ratatoskr} are the most popular. Booksim \cite{booksim} uses a channel model that implements a two-phase evaluate-update protocol for routers. Noxim \cite{noxim} is implemented using the cycle-accurate modeling of the SystemC library. It allows varying NoC parameters with configuration files. Ratatoskr \cite{Ratatoskr} also builds upon SystemC but uses TLM for system-level benchmarks. A 64-node mesh network yields a simulation frequency between 1000 Hz and 10 000 Hz (cf. Fig. \ref{fig:sim_perf}). 

For even higher speed, emulation on FPGAs became the industry standard. 
A whole NoC might be too large for a single FPGA, so Kouadri et al. \cite{kouadri2007scalable,kouadri2008large} emulate it through partitioning the NoC and mapping it onto multiple FPGAs. However, the measurement accuracy is limited by the off-chip data transmission architecture. When using only a single FPGA, one can directly map the NoC (if the FPGA is large enough) or use time-division multiplexing (TDM). TDM \cite{wolkotte2007fast,papamichael2011fast,pellauer2011hasim} implements a single router in the FPGA's programmable logic.  
The status of the routers are stored in the off-chip memory. 
This method can emulate a large-scale NoC with over 1000 nodes, but it requires huge off-chip memory and deteriorates performance. Chu et al. \cite{chu2015ultra,chu2017fast} partition the NoC into multiple virtual clusters, each cluster containing multiple routers. Each cluster is emulated sequentially, increasing emulation speed. 3D NoCs can also be implemented using TDM \cite{dhoore3dNoC}.

\respond{TDM is rarely used in industry, since FPGAs can easily be clustered to provide sufficient resources to accommodate the whole NoC. In other words, directly-mapped (DM)} approaches became viable. This method can achieve the highest performance because all routers run in parallel. Different framework designs will degrade the emulated NoC frequency.
AcENoC \cite{acenocs2011} ran a 5$\times$5 NoC using a separate bus system to transmit the data to/from the NoC and achieved a maximum frequency of 23kHz. For larger 8$\times$8 NoCs, Drewes et al. \cite{drewes2017fpga} used an AXI4-Stream bus to collect the data from the NoC with the technique of an asynchronous serializer, and achieved 16kHz.

To further boost performance, Netrace's dependency tracking between packets was implemented in hardware by \cite{chu.2020} achieving up to 12MHz emulation speed. However, this limits the system's flexibility, as the benchmark cannot be replaced easily. 

A complete comparison of all NoC emulation systems is given in Tab. \ref{tab:cmp_emu}. Only \systemname\ provides the flexibility for changing applications at high frequency.

%% file: tables/cmp.tex
\newcommand{\cmark}{\textcolor{green}{\ding{51}}}%
\newcommand{\xmark}{\textcolor{red}{\ding{55}}}%

\begin{tabular}{lcccccc}

    \toprule
    Emulator & Model & TDM/DM & Transactor & Synthetic & \multicolumn{2}{c}{Real benchmark} \\
    &&&& benchmark & Netrace \cite{netrace2010} & Edge-AI \\

    \midrule

    Kouadri \cite{kouadri2007scalable,kouadri2008large} & Pure HW & - & - & \cmark & \xmark & \xmark \\
    
    Wolkotte \cite{wolkotte2007fast} & Hybrid & TDM & Bus & \cmark & \xmark & \xmark \\
    
    Papamichael \cite{papamichael2011fast} & Hybrid & TDM & Bus & \cmark & \xmark & \xmark \\
    
    HAsim \cite{pellauer2011hasim} & Hybrid & TDM & Bus & \cmark & \xmark & \xmark \\
    
    FNoC \cite{chu2017fast,chu2015ultra} & Pure HW & TDM & - & \cmark & \xmark & \xmark \\
    
    D'Hoore \cite{dhoore3dNoC} & Pure HW & TDM & - & \cmark & \xmark & \xmark \\
    
    Chu \cite{chu.2020} & Pure HW & TDM & - & \xmark & \cmark & \xmark \\
    
    AcENoCs \cite{acenocs2011} & Hybrid & DM & Bus & \cmark & \xmark & \xmark \\
    
    Drewes et al. \cite{drewes2017fpga} & Hybrid & DM & Bus+Stream & \cmark & \cmark & \xmark \\
    
    \systemname & Hybrid & DM & Stream & \cmark & \cmark & \cmark \\

    \bottomrule

\end{tabular}

%% file: method.tex
The previous state-of-the-art emulation frameworks using directly-mapped NoCs are limited in performance from a) the bus system data transaction \cite{drewes2017fpga,acenocs2011}, and b) the software clock halting technique for cycle-accurate emulation \cite{drewes2017fpga}.
This paper presents an improved approach tackling both downsides to achieve an even faster emulation speed. 

\begin{figure}
    \centering
    \includegraphics[width=1\linewidth]{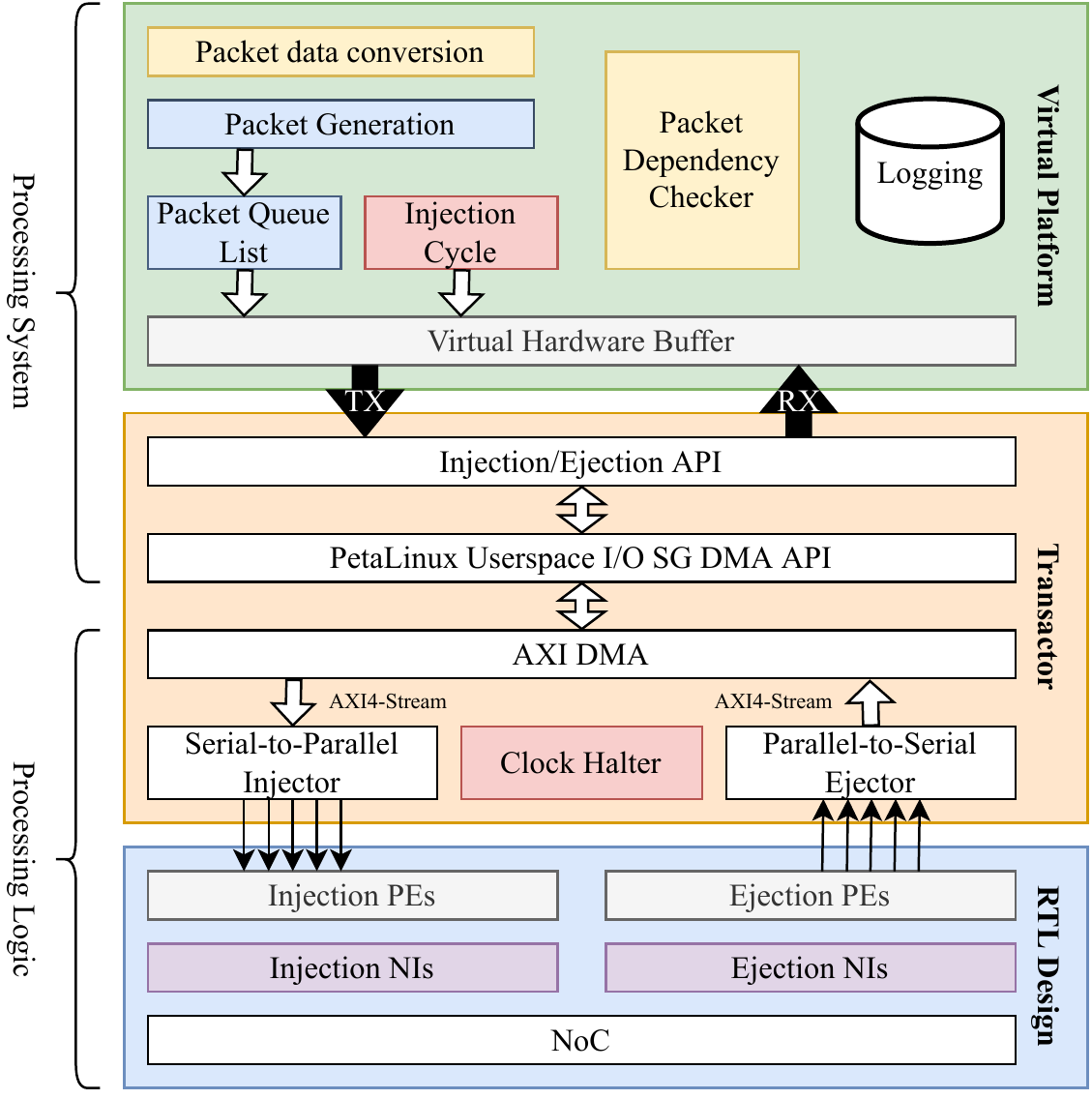}
    \caption{\systemname\ system architecture.}
    \label{fig:coemu_sys}
\end{figure}

The architecture of \systemname\ is shown in Fig. \ref{fig:coemu_sys}. It consists of the software (virtual platform/green), executed in the CPU cores of our FPGA, the hardware (RTL design/blue), executed in the programmable logic, and a transactor (orange) that connects both of them. 
We propose a novel transactor to overcome the performance bottleneck with the AXI4-Stream data transaction and hardware-based clock halting technique.
On the software side, a virtual platform generates and sends packets to the RTL model at a defined time quantum (\emph{injection cycle}) through the transactor and places it in the virtual hardware buffer. A clock halter enables to stop the execution of the RTL model at any given time. This unit is used in two scenarios: 
For packet injection, the transactor sets the time quantum step to the clock halter through the serial-to-parallel injector and enables RTL execution until the next given \emph{injection cycle}. When the NoC RTL model is executed (\ie not halted), the serial-to-parallel injector injects packets into the source PEs.
The parallel-to-serial ejector halts the RTL model via the clock halter when the packets arrive at a destination processing element (PE) for packet ejection. Then, it sends the halted time step (recorded by the clock halter) and the arrived packets via DMA to the software-side buffer through the transactor. The virtual platform checks, removes, and logs received packets.

\subsection{Hardware Architecture}
\begin{figure*}
    \centering
    \includegraphics[width=.95\textwidth]{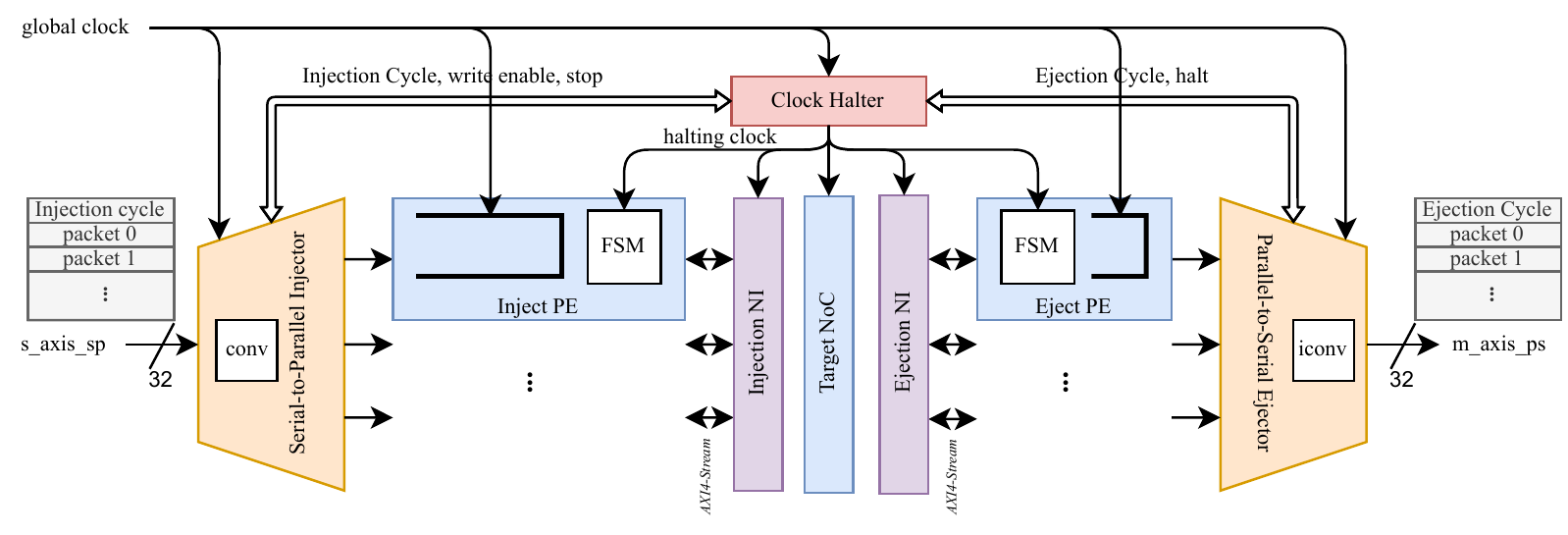}
    \caption{\systemname\ hardware architecture.}
    \label{fig:hw_noc_validation}
\end{figure*}

Fig. \ref{fig:hw_noc_validation} shows the hardware architecture of our novel transactor with its adjacent units.

\subsubsection{Clock Halter}

\begin{figure}[h]
    \centering
    \includegraphics[width=.9\linewidth]{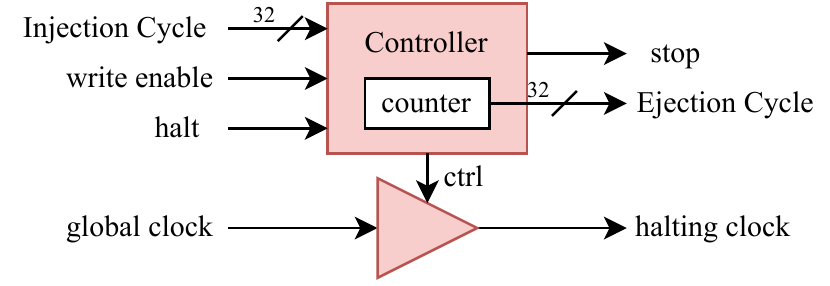}
    \caption{Block diagram of the clock halter.}
    \label{fig:clock_halter}
\end{figure}

The clock halter stops the execution of the emulated NoC at any time for synchronization with the virtual platform. It is a central logic connected to nearly all other components (see Fig. \ref{fig:hw_noc_validation}). Fig. \ref{fig:clock_halter} shows its block diagram.
The \emph{halting clock} is generated by a buffer driven by the \emph{global clock} and enabled by a \emph{ctrl} signal (=1).
An \emph{injection cycle} is stored using the \emph{write enable} signal.
The \emph{counter} counts cycles. The \emph{ctrl} signal enables the \emph{halting clock} when the value is smaller than the stored \emph{injection cycle} value.
The signal \emph{ctrl} halts the buffer and the \emph{counter} stops counting whenever the \emph{halt} signal is 1.
If the \emph{counter} equals to the \emph{injection cycle}, the \emph{stop} signal is set and the \emph{halting clock} is disabled.

\subsubsection{Serial-to-Parallel Injector}

The serial-to-parallel injector receives packets from the software side and injects them into the NoC. For communication with the software, the injector contains an AXI4-Stream slave port. 
When a transaction starts, the first stream data determines the \emph{injection cycle} until which the RTL design is executed. Once reached, the injector converts the packet into the header flits (the communication unit within the NoC; \emph{conv} in Fig. \ref{fig:hw_noc_validation}) and sends them to the respective PE's FIFO using the packet's source address.

\subsubsection{Parallel-to-Serial Ejector}
\begin{figure}[h]
    \centering
    \includegraphics[width=.85\linewidth]{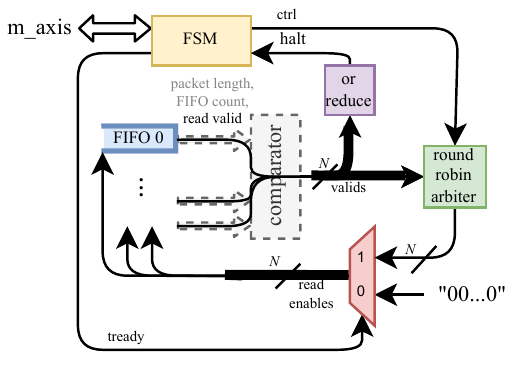}
    \caption{Serializer as used in the ejector; gray-shaded part for multi-VC NoCs.}
    \label{fig:sync_cmp_serializer}
\end{figure}

When a destination PEs receives a complete packet (\ie all of its flits), the parallel-to-serial ejector instructs the clock halter to halt the RTL design through the \emph{halt} signal (=1).
This module converts the header flit back (\emph{iconv} in Fig. \ref{fig:hw_noc_validation}) to packet data and sends it via DMA (AXI4-Stream) to the software side.
The sent data contain the clock cycle at which the packet was received. 
Fig. \ref{fig:sync_cmp_serializer} shows the implementation logic of the single clock serializer.
The blue part refers to the corresponding PE's FIFO.
If a complete packet arrives at the destination, only the header flit is stored in the FIFO, the corresponding signal \emph{read valid} becomes 1.
The block \emph{or reduce} (purple) tells the FSM (yellow) to initiate the stream transaction.
The FSM updates the \emph{round robin arbiter} (green) with the signal \emph{ctrl} (=1) to decide which FIFO to read.
When data is read from the AXI4-stream port (\emph{tready}=1), the corresponding FIFO is read through the multiplexer's output signal (red).
If all header flits in the FIFOs are ejected (\emph{halt}=0), the round-robin arbiter (selecting the ejection FIFO order) will be updated, and the RTL emulation will continue.

\subsubsection{Injection PE}
The injection PE subsequently injects flits of each packet into the NoC. For each PE, there is a network interface (NI) that handles the assignment of flits to VCs and sends them into the connected router (see below). 

The injection PE contains a FIFO and an FSM, managing the AXI4-Stream transactions and packet injection via a NI.
When the PE's FIFO holds the first flits of a packet (header flit), it starts the transaction by injecting it. The packet contains dummy payload flits as our implementation does not transmit "useful" payloads and handles the packet assignment at the destination via the software's virtual buffers.

\subsubsection{Injection NI}
The NI accepts a complete packet in one AXI4-Stream transaction.
If the NoC uses multiple VCs, packets will be assigned via round-robin arbitration. The flits travel through the NoC until they are received in the target PE's NI. In the implementation at hand, we use the Ratatoskr router \cite{joseph2019nocs}; it can be exchanged by any other NoC that implements an AXI4-compatible interface.

\subsubsection{Ejection NI}
The ejection NI contains one FIFO per VC; the FIFO is long enough to store a whole packet, \ie all of its flits.
When all packet flits are received, the NI starts the AXI4-Stream transaction and sends the packet to its PE. The number of flits is checked via a comparator in Fig. \ref{fig:sync_cmp_serializer}.

\subsubsection{Ejection PE}
This PE functions like the injection PE. It receives flits from its NI. Only the head flit is stored to be transmitted to the software side.
This flit is put into a 1-flit FIFO connected to the serializer when a packet is completed.

\subsection{Software Architecture}

Drewes et al. \cite{drewes2017fpga} used the simple DMA mode to transfer data to the NoC.
\systemname\ uses Scatter Gather (SG) DMA mode to improve performance. On the software side, the SG DMA API for PetaLinux Userspace I/O \cite{plnxuio} is implemented (Fig. \ref{fig:coemu_sys}).
We also used compiler optimizations, \eg \texttt{-Ofast}.

\begin{figure*}
    \centering
    \includegraphics[width=\textwidth]{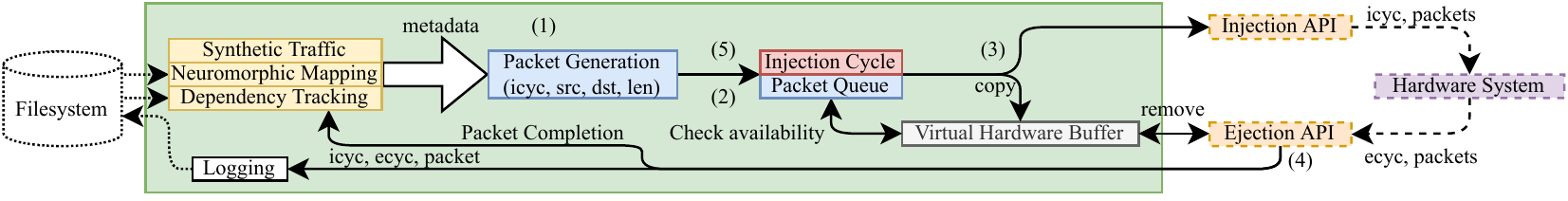}
    \caption{\systemname\ software architecture.}
    \label{fig:sw_sys}
\end{figure*}

Fig. \ref{fig:sw_sys} shows the software design of the framework. It is the detailed view of the virtual platform in Fig. \ref{fig:coemu_sys}.
The variables \emph{icyc} is the \emph{injection cycle}, \emph{src} the packet's source address, \emph{dst} the destination address, and \emph{len} the packet length/flit count.

In general, our software is executed in the following six steps (highlighted in Fig. \ref{fig:sw_sys}):

\begin{enumerate}[label=(\arabic*)]
\item Packet data is generated using the flexible software interface (example see below).

\item The program searches for the earliest available packets and puts them into the queue.

\item The \emph{virtual hardware buffer} sends the packet data to the Injection PE's FIFO. It also keeps a copy to handle the assignment of the received flits to the correct packet, as explained above.
The \emph{injection cycle} and the packets are sent to the NoC and transmitted there to the target PE.

\item When the packets have arrived at their destination, the DMA stored them in the main memory.
Then, the program compares the received packet, which is matched with its counterpart in the \emph{virtual hardware buffer} to enable dependency tracking.

\item After injecting the previous time quantum in (3), the program needs to determine the next time quantum to inject the packets. Then, the program checks whether the next time quantum has exceeded the user-defined maximum cycle to run.
If the program reaches the maximum cycle or no more packet to inject, it goes to (6); else to (2).

\item If the virtual buffer is empty, the emulation will stop.
\end{enumerate}

We will demonstrate the flexibility of this software architecture with three different traffic scenarios (see below).
Any other use case can be implemented easily by modifying the software code.
A simple example code exemplifies this (Listing \ref{lst:sw}).
Line 1 refers to the yellow blocks in Fig. \ref{fig:sw_sys}, different modes are decided, and the metadata is generated.
The rest of the program refers to \respond{the six steps}, in which a packet list is iterated and injected to the NoC.

\begin{lstlisting}[label={lst:sw}, style=CStyle, caption={Example packet generation.}]
metadata = initialization(mode);
pkt_cyc_list = generate_packets(metadata);
do
{
    if (cyc < max_cyc)
        put_packet_to_queue(cyc, pkt_cyc_list, queue_lists);
    hw_list = copy_to_hw_buffers_and_create_hw_list(hw_buffers, queue_lists);

    inject(cyc, hw_list);
    eject(hw_buffers);

    cyc = calculate_next_injection_cycle(pkt_cyc_list);
} while (cyc < max_cyc);
check_lost_packets(hw_buffers);
\end{lstlisting}

%% file: result.tex
\subsection{Hardware Costs}

Table \ref{tab:hw_resc} shows the used FPGA resources of \systemname, AcENoC \cite{acenocs2011}, Drewes et al. \cite{drewes2017fpga} and Chu \cite{chu.2020}.
\systemname's resources were obtained from Vivado 2018.2 for a Zynq UltraScale+ MPSoC ZCU102.
The \emph{global clock} (Fig. \ref{fig:hw_noc_validation}) is set to 80MHz and the FIFOs in the NoC and the transactor use the Xilinx's FIFO IP \cite{xlnxFIFO2017} to support larger setups.

As we can see, the FPGA resources of \systemname\ increase approximately linearly with the router count.
\systemname\ used more memory (LUTRAM, BRAM) than AcENoC \cite{acenocs2011} and Drewes et al. \cite{drewes2017fpga}, which is required for our better emulation performance. Still, we can host up to 169 routers, more than triple the router count than in the previous \respond{DM} framework (even with larger single routers). We achieved this by enabling them to use larger standard FPGAs;  \cite{drewes2017fpga} relied on a custom clock halter that only was possible in their FPGA. Chu \cite{chu.2020} consumes more resources than \systemname\ (8$\times$8) because they implement the dependency tracking in hardware.

\begin{table}
    \centering
    \caption{Hardware resources of different NoC configurations of each emulator. (FB: flit buffer)}
    \resizebox{0.5\textwidth}{!}{\input{./tables/hw_resc.tex}}
    \label{tab:hw_resc}
\end{table}

\subsection{Emulation Performance}
\systemname\ is validated with uniform random traffic (uniform random source-destination pairs and injection times). \respond{This traffic allows to fuzzzy-test the NoC as random traffic is sent though the network.}

\begin{figure}
    \centering
    \includegraphics[width=\linewidth]{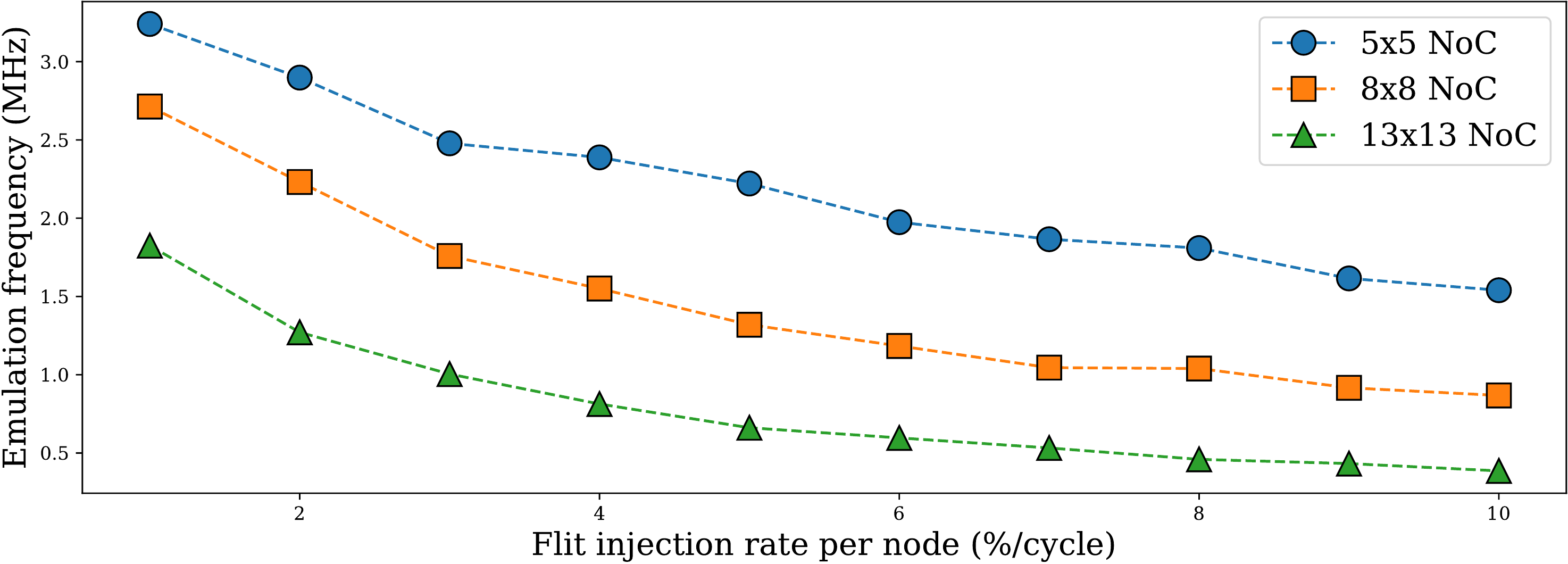}
    \caption{Emulation performance with uniform random traffic.}
    \label{fig:uperf}
\end{figure}

To compare \systemname's performance with the state-of-the-art emulation, the NoC is set to the same configuration as AcENoC \cite{acenocs2011} (5$\times$5 mesh with 2 VCs and 8-flit buffer) and Drewes et al. \cite{drewes2017fpga} (8$\times$8 mesh with 2 VCs and 3-flit buffer).
The largest NoC that can be emulated on our FPGA is 13$\times$13 mesh NoC with 2 VCs and 4-flit buffer.
The emulation performance of these configurations are shown in Fig. \ref{fig:uperf}. We observe a performance degradation with NoC size and injection rate.

Table \ref{tab:perf} shows the emulation frequency at 5\% flit injection rate for synthetic traffic. \systemname\ achieves 2221 kHz for 5$\times$5 mesh, 1319 kHz for 8$\times$8 mesh.
\systemname\ yields a 96.6$\times$ speedup over AcENoC \cite{acenocs2011}, and a 79.3$\times$ speedup over Drewes et al. \cite{drewes2017fpga}.
We have achieved faster emulation speed than any other flexible, directly-mapped framework.

\subsection{Performance Scaling}

As stated, the emulation performance drops with NoC size and traffic load. We compare our system against simulators to analyze this scaling effect as they show the same behavior from the similar root of traffic injection. 

\begin{figure}
    \centering
    
    \subfigure[Simulation performance for different flit injection rates in a 13$\times$13 NoC.]{\includegraphics[width=\linewidth]{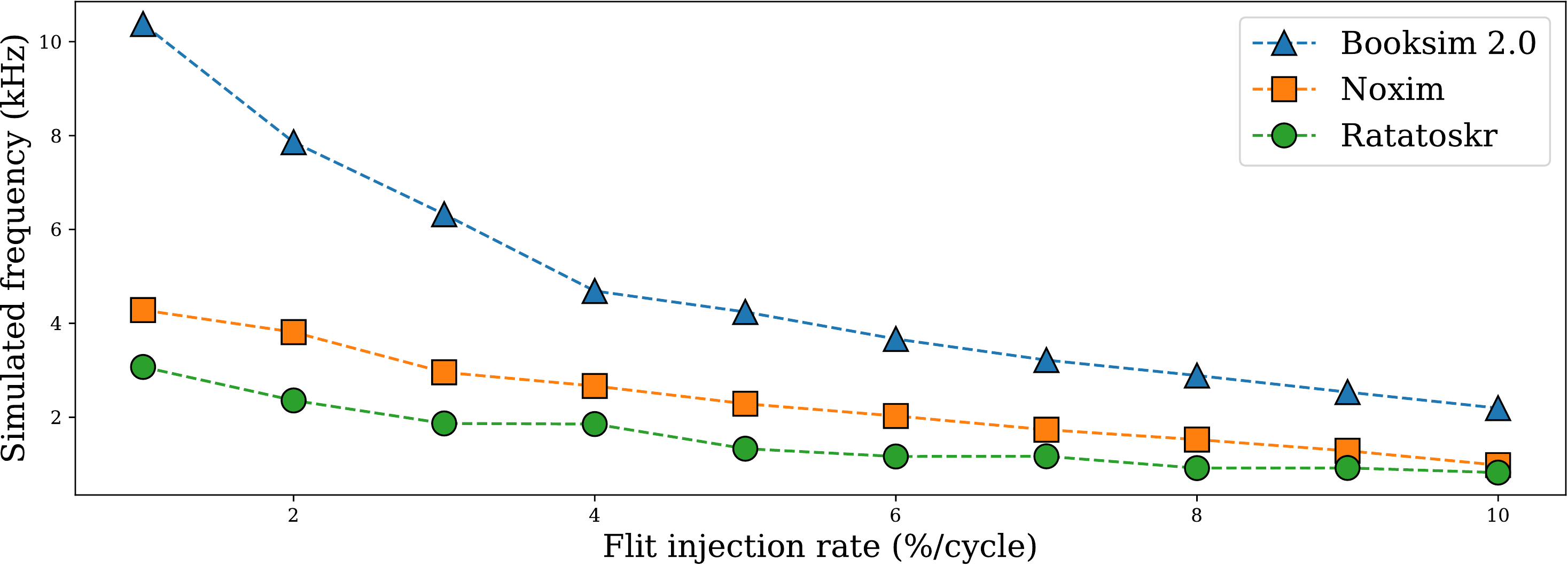}}
    
    \subfigure[Simulation performance for different NoC sizes at 0.5\% flit injection rate.]{\includegraphics[width=\linewidth]{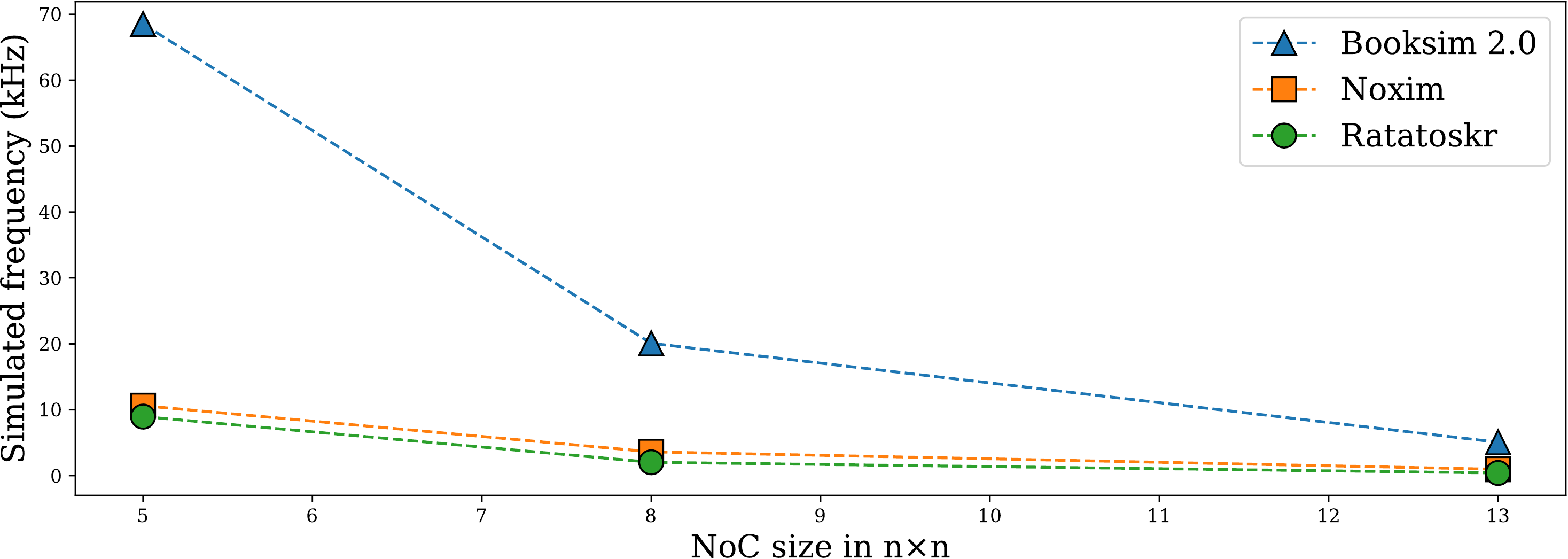}}

    \caption{Simulation performance with NoCs.}

    \label{fig:sim_perf}

\end{figure}

We simulate a 13$\times13$ mesh NoC (2 VCs and 4-flit buffer) with the simulators Booksim 2.0 \cite{booksim}, Noxim \cite{noxim}, and Ratatoskr \cite{Ratatoskr} using dimension-ordered routing, 5-flit packets. We measure the median of 10 simulations on a WSL Ubuntu 20.04.2 LTS using one Intel i7-5700HQ core at 2.7 GHz and show the results in Fig. \ref{fig:sim_perf}.
By increasing the flit injection rate, we can see the simulation performance decreases (Fig. \ref{fig:sim_perf}(a)).
Fig. \ref{fig:sim_perf}(b) shows three different NoC sizes at fixed 5-\% injection rate.
A larger injection rate or NoC size degrades the simulation performance because the simulation needs to generate traffic and simulate the routers sequentially. 

Analyzing the performance drop of these simulators and \systemname, we found out that Booksim 2.0 \cite{booksim} has the highest performance loss from 1 to 10\% injection rate for 13$\times$13 NoC (78.9\%; \systemname\ is 78.8\%; Noxim \cite{noxim} is 77.1\%; Ratatoskr \cite{Ratatoskr} is 73.3\%). Emulation behaves similarly to the simulations because of the software-side traffic generation.

If the NoC \respond{size is increases from} 5$\times$5 to 13$\times$13 (fix 5\% flit injection rate per cycle), Ratatoskr \cite{Ratatoskr} has  lost the most performance (95.4\%) compared to Booksim 2.0 \cite{booksim} (92.6\%), Noxim \cite{noxim} (90.8\%) and \systemname\ (70.2\%).
Here we observe an advantage of emulation, as \systemname\ yields the lowest performance drop.

\begin{table}
    \centering
    \caption{Performance comparison of frameworks.}
    \input{./tables/perf.tex}
    \label{tab:perf}
\end{table}

\subsection{Case Study I: Multi-core Processors}

\begin{figure}
    \centering
    \includegraphics[width=\linewidth]{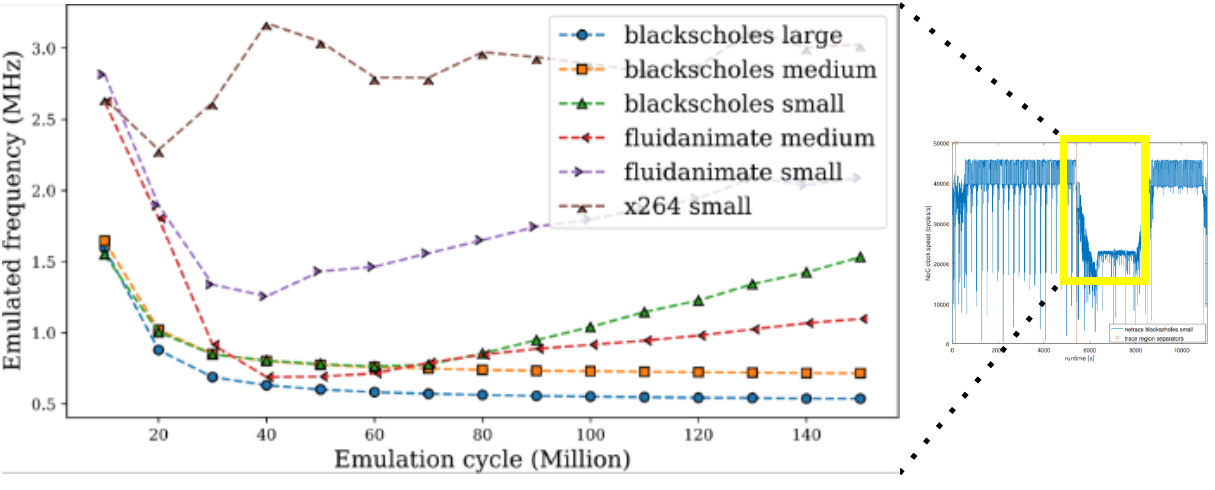}
    \caption{Emulation performance for PARSEC \cite{netrace2010} ROI for different traces. (8$\times$8 NoC with 2 VCs, 3-flit buffer). The  right-hand-side image is cited from \cite{drewes2017fpga}.}
    \label{fig:cmp_nperf}
\end{figure}

Netrace \cite{netrace2010} provides 64-core processor's traces with dependency tracking. It contains five phases from the PARSEC benchmarks: the startup, warmup, region of interest (ROI), result output, and post benchmark.
Drewes et al. \cite{drewes2017fpga} has run the whole benchmark (Fig. \ref{fig:cmp_nperf} right-side yellow box). Their results show a performance drop in the ROI (yellow highlighted) because this region contains the highest traffic workload.
As the ROI is hence the exciting part, only it is investigated in our experiment. The result is shown in Fig. \ref{fig:cmp_nperf}, left-hand side.
Similar to \cite{drewes2017fpga}, we first observe a performance drop then followed by a performance recovery.
In average, we achieve 1426 kHz (see Table \ref{tab:perf}), which has achieved $36.3\times$ speedup compared to Drewes et al. \cite{drewes2017fpga}.
Our framework is slower than Chu \cite{chu.2020} by 0.11$\times$. The reason is that Chu \cite{chu.2020} implemented Netrace-specific dependency-tracking hardware. \respond{While this method has a high performance, it is not flexible for application-centered engineers to adopt different use cases easily. Therefore, our software-based dependency tracking offers a compelling trade-off between performance and flexibility for many practitioners.}

\begin{figure}
    \centering

    \subfigure[1 VC, 2-flit buffer]{\includegraphics[width=.34\textwidth]{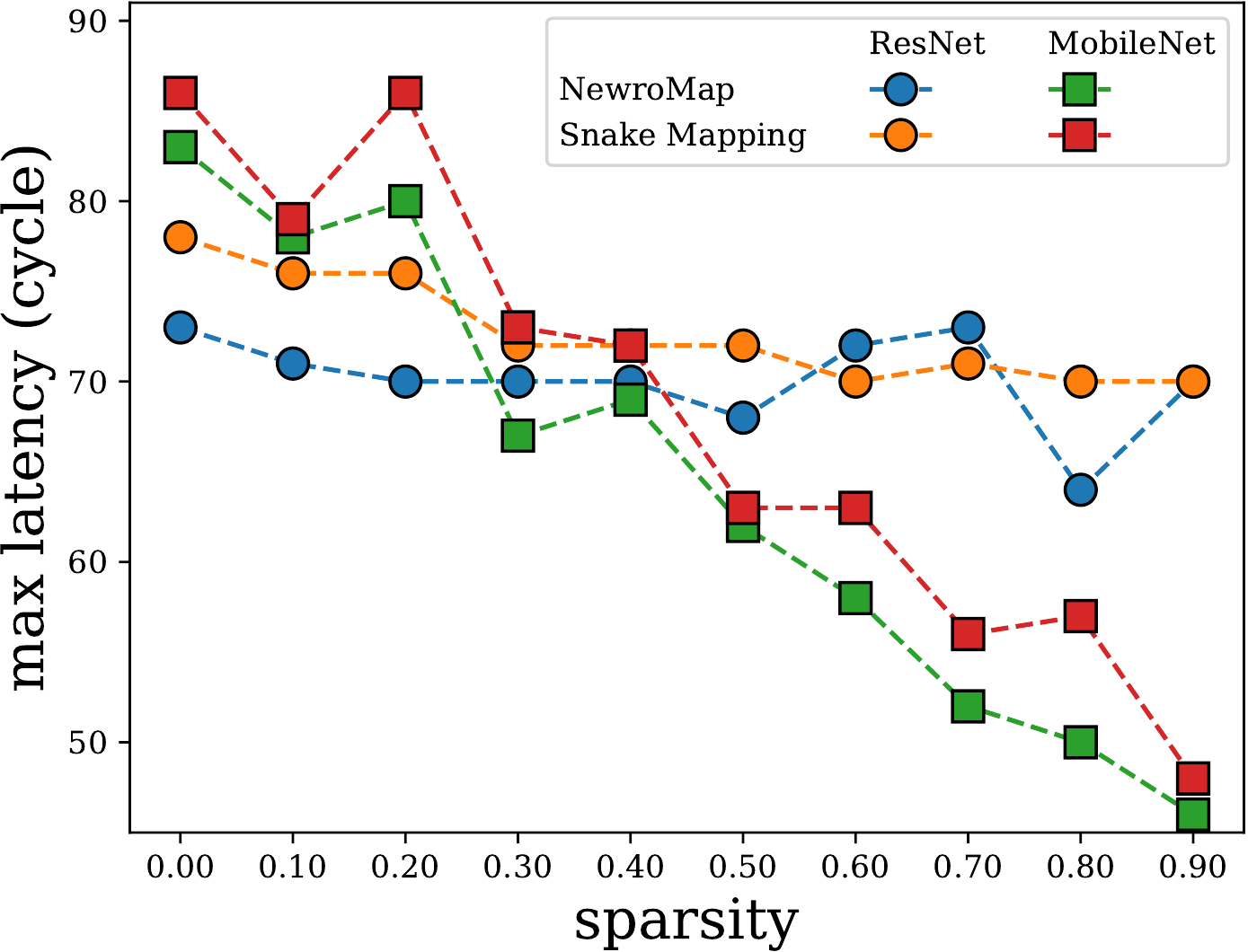}}\\%
    \subfigure[2 VCs, 1-flit buffer]{\includegraphics[width=.34\textwidth]{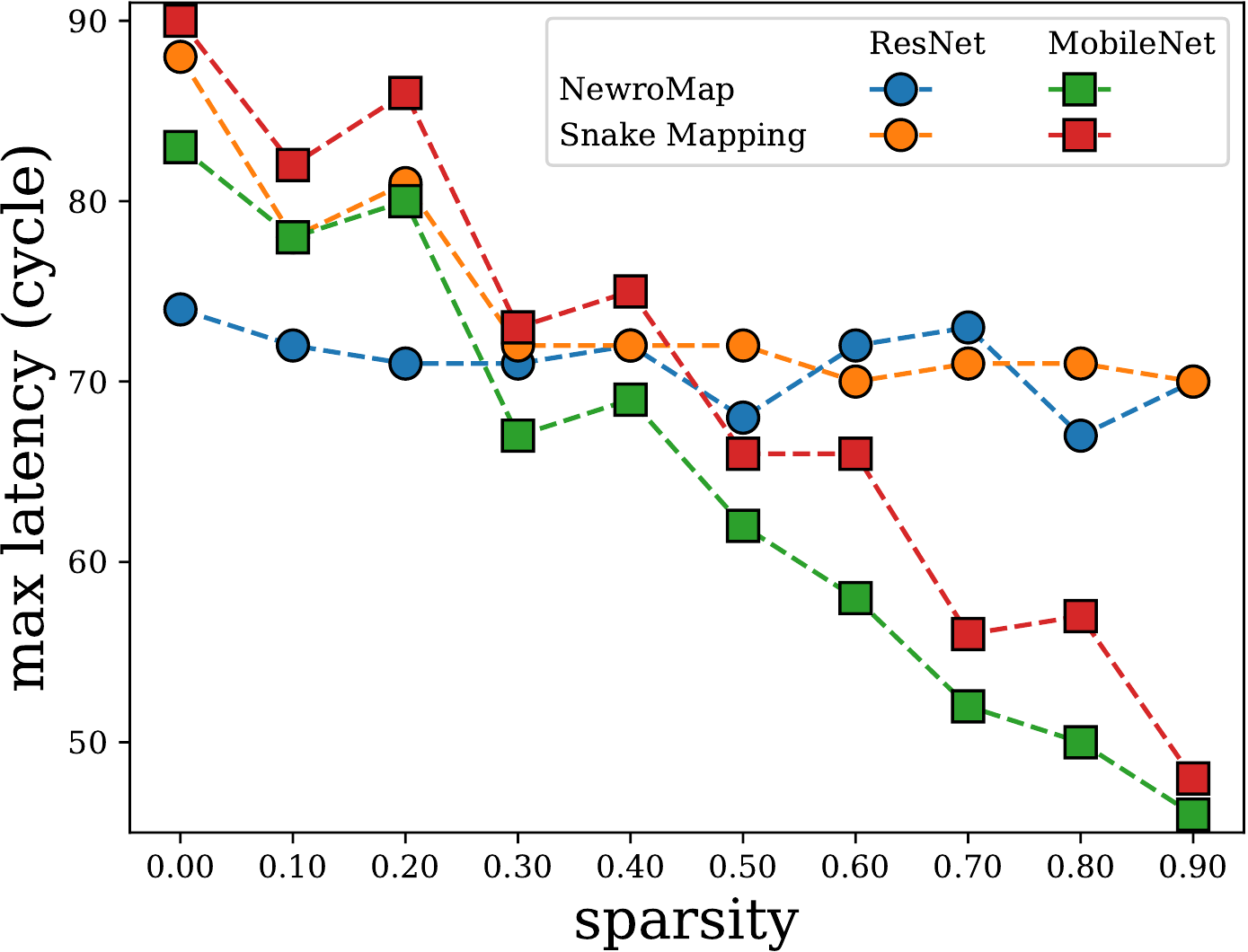}}\\%
    \subfigure[2 VCs, 2-flit buffer]{\includegraphics[width=.34\textwidth]{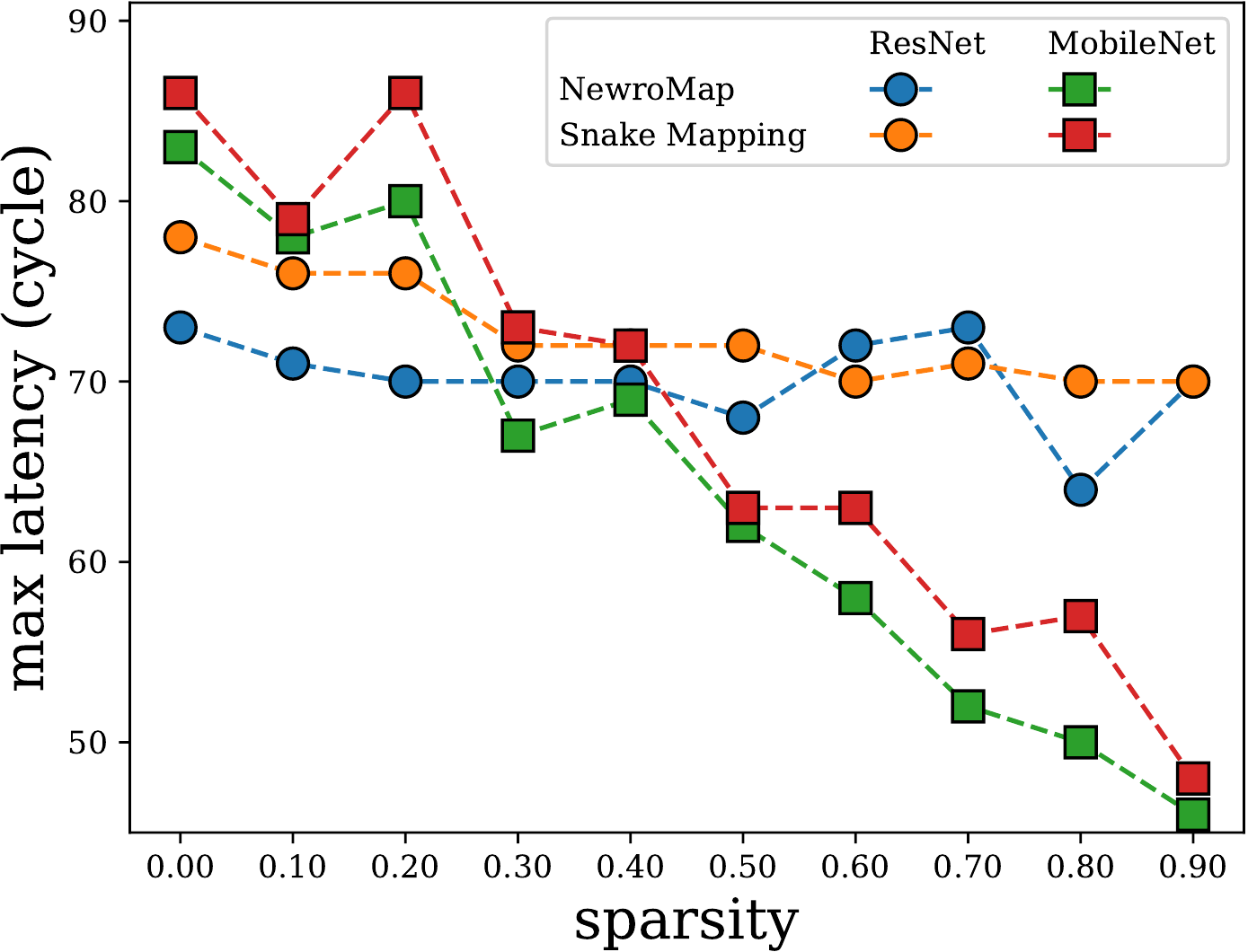}}%
\vspace{-1ex}
    \caption{Maximum latency measurement for different CNN mappings \cite{newromap} for sparsity rates.}
    \label{fig:lat_vs_spar}

 \end{figure}

\subsection{Case Study II: Neuromorphic Edge-AI Accelerator}

Studying edge AI-accelerator architectures \cite{ISAAC,puma,chi2016prime,liu2015reno} shows that all of them demand a scalable NoC.
This yields a workload mapping problem, \eg solved by NewroMap \cite{newromap} for CNNs, in which the network activations are transmitted via the NoC. 

One well-known property of neural networks is their high sparsity, \ie\ 0-values, \respond{which need not be sent} via the NoC \cite{neunoc2018}. Hence, the effective injection rate in the NoC for each mapping is scaled by a) a sparsity factor and b) the target framerate in our exemplary case of video applications. This gives the formula for the injection per PE rate as $$irate = \frac{map_{neurons}*(1-sparsity)*framerate}{frequency_{NoC}},$$ where $map_{neurons}$ is the number of neurons mapped to a core. We use the $framerate$ and $frequency_{NoC}$ from one commercially available accelerator called NeuronFlow \cite{neuronflow} as 30 FPS and 1 GHz.


We analyzed different mappings and NoC architectures to demonstrate the flexibility of \systemname. We plot the maximum packet latency in Fig. \ref{fig:lat_vs_spar}. We used very lightweight NoCs with/without VCs and multi-flit buffers. As edge AI workloads tend to have particular traffic patterns with high locality, these architectures are promising to investigate (even though they would be not useful for conventional multi-core CPUs).

We observe that the maximum packet latency decreases with higher sparsity for all architectures. This effect is as expected as less traffic yields less congestion. We further observe that the optimized mappings proposed by NewroMap can improve latency vs. snake mapping, which was previously only demonstrated in simulations in \cite{newromap}.

Another interesting finding is that a NoC without VCs and 2-flit buffers has a lower maximum latency than the architectures with 2 VCs and 1-flit buffers (see Fig. \ref{fig:lat_vs_spar}(a) vs.\ Fig. \ref{fig:lat_vs_spar}(b)). Both architectures yield approximately the same area costs. At first glance, this is surprising since VCs promise better peak performance. However, the high locality of edge-AI traffic patterns effectively removes the need for VCs in many routers. Also, adding additional buffers (Fig. \ref{fig:lat_vs_spar}(c)) does not improve performance. These findings advocate for very lightweight NoCs in edge-AI systems. However, the authors would like to mention that a VC-less router will not be the best architecture to choose in all cases. While it has the lowest area costs and best performance, it effectively prohibits multi-thread processing of CNN layers on single cores. Hence, the higher NoC costs might be worth it from a system design perspective.

%% file: tables/hw_resc.tex
\begin{tabular}{*{10}{c}}

    \toprule
    
    Emulator & NoC & VC & FB &	LUT & LUTRAM & FF & BRAM & BUFG & Slices \\
    
    \midrule
    
    \systemname & 5$\times$5 & 2 & 8 & 40750 & 6150  & 45944  & 39.5  & 3 & -  \\
    \systemname & 8$\times$8 & 2 & 3 & 100148  & 14934  & 105960  & 98  & 3 & - \\
    \systemname & 13$\times$13 & 2 & 4 & 267230  & 39174  & 274644  & 255.5  & 3 & - \\
    \systemname & 13$\times$13 & - & - & 34261 &	1110 &	38666 &	86.5 &	3  & -\\
    
    AcENoC \cite{acenocs2011} & 5$\times$5 & 2 & 8 & 52520 & 840 & 19569 & - & - & -\\
    Drewes \cite{drewes2017fpga} &  8$\times$8 & 2 & 3 & 97515 & - & 78361 & 36.5 & - & -\\
    
    Chu \cite{chu.2020} & 8$\times$8 &- &- & 133327 & - & 111576 & 830 & - & 43903 \\

    \bottomrule


\end{tabular}

%% file: tables/perf.tex
\begin{tabular}{*{5}{c}}

    \toprule
    Emulator & NoC & Traffic & Frequency & Speedup \\
    
    & & & [kHz] & \\

    \midrule
    
    AcENoC \cite{acenocs2011}           & 5$\times$5 & Synthetic & 23     & 96.6 \\
    Drewes et al. \cite{drewes2017fpga} & 8$\times$8 & Synthetic & 16.639 & 79.3 \\
    Drewes et al. \cite{drewes2017fpga} & 8$\times$8 & Netrace   & 39.243 & 36.3  \\
    Chu \cite{chu.2020}                 & 8$\times$8 & Netrace   & 12979 & 0.11 \\

    \systemname & 5$\times$5 & Synthetic & 2221.464 & - \\
    \systemname & 8$\times$8 & Synthetic & 1319.333 & - \\
    \systemname & 8$\times$8 & Netrace   & 1426.404 & - \\
    
    \systemname & 13$\times$13 & Synthetic & 661.291 & - \\

    \bottomrule

\end{tabular}

%% file: con.tex
This paper proposed a fast and flexible FPGA-based NoC hybrid emulation called \systemname. These features are achieved by a novel transactor architecture and a programmable software interface. The transactor contains a novel clock-synchronization method and hardware-only packet generation unit. The programmable software interface enables mapping different system benchmarks to the NoC, which is highly relevant as NoCs are widely used today. \systemname\ has achieved 36$\times$ to 96$\times$ speedup compared to the comparable previous \respond{DM method}. \respond{We also increased the area efficiency and were able to emulate an NoC with 169 routers on a single FPGA with a state-of-the-art size at time of writing this paper.}
We used the emulator in two case studies to demonstrate its practical use for architects.